\documentclass[12pt]{iopart}
\usepackage{bm}
\usepackage{graphicx}
\usepackage{color}

\begin{document}

\title{Electronic properties in itinerant ferromagnet SrRu$_{1-x}$Ti$_x$O$_3$}

\author{Renu Gupta$^1$, R. Rawat$^2$ and A K Pramanik$^1$}
\address{$^1$School of Physical Sciences, Jawaharlal Nehru University, New Delhi - 110067, India}
\address{$^2$UGC-DAE Consortium for Scientific Research, University Campus, Khandwa Road, Indore 452001, India}

\eads{\mailto{akpramanik@mail.jnu.ac.in}}

\begin{abstract}
Here, we study the electrical transport and specific heat in 4$d$ based ferromagnetic material SrRuO$_3$ and its Ti substituted SrRu$_{1-x}$Ti$_x$O$_3$ series ($x$ $\le$ 0.7). The SrRuO$_3$ is a metal and shows itinerant ferromagnetism with transition temperature $T_c$ $\sim$ 160 K. The nonmagnetic Ti$^{4+}$ (3$d^0$) substitution would not only weaken the active Ru-O-Ru channel but is also expected to tune the electronic density and electron correlation effect. A metal to insulator transition has been observed around $x$ $\sim$ 0.4. The nature of charge transport in paramagnetic-metallic state ($x$ $\leq$ 0.4) and in insulating state ($x$ $>$ 0.4) follows modified Mott's variable range hopping model. In ferromagnetic-metallic state, resistivity shows a $T^2$ dependence below $T_c$ which though modifies to $T^{3/2}$ dependence at low temperature. In Ti substituted samples, temperature range for $T^{3/2}$ dependence extends to higher temperature. Interestingly, this $T^{3/2}$ dependence dominates in whole ferromagnetic regime in presence of magnetic field. This evolution of electronic transport behavior can be explained within the framework of Fermi liquid theory and electron-magnon scattering mechanism. The negative magnetoresistance exhibits a hysteresis and a crossover between negative and positive value with magnetic field which is connected with magnetic behavior in series. The decreasing electronic coefficient of specific heat with $x$ supports the increasing insulating behavior in present series. We calculate a high Kadowaki-Woods ratio ($x$ $\leq$ 0.3) for SrRuO$_3$ which increases with substitution concentration. This signifies an increasing electronic correlation effect with substitution concentration.
\end{abstract}


\submitto{\JPCM}

\maketitle
\section {Introduction}
Due to its complex magnetic and transport properties, the 4$d$ based oxide SrRuO$_3$ continue to attract attention where many of the observed properties are important for technological applications.\cite{koster,bensch,noro} This material is commonly believed to an itinerant ferromagnet (FM) with a transition temperature $T_c$ $\sim$ 160 K.\cite{koster,cao,fuchs,cheng,allen,mazin} SrRuO$_3$ further exhibits a metallic behavior where the resistivity ($\rho$) shows a linear increase till temperature as high as 1000 K, even crossing the Ioffe-Regel limit which is considered as limit for good metallic conduction.\cite{allen} The electrical charge conduction, however, has direct correlation with the magnetic ordering in this material as $\rho(T)$ shows a distinct slope change across $T_c$ where its value decreases with faster rate with decreasing temperature. While photo-emission spectroscopy indicates a weak or moderate electron correlation strength ($U$) in SrRuO$_3$, \cite{fujioka,okamoto,takizawa} this material shows a reasonably high electronic coefficient of specific heat $\gamma$ ($\sim$ 30 mJ mol$^{-1}$ K$^{-2}$) and a Fermi-liquid (FL) behavior at low temperature. \cite{allen,cao}       

To understand its exotic magnetic as well as transport properties, the chemical substitution with suitable dopant character at Ru-site has often been used. In present study, we have used nonmagnetic Ti$^{4+}$ (3$d^0$) substitution to replace magnetic Ru$^{4+}$ (4$d^4$). While a least structural modification is expected due their matching ionic radii (Ru$^{4+}$ = 0.62 \AA and Ti$^{4+}$ 0.605 \AA), but this substitution is expected to introduce random potential through site disorder in active Ru-O-Ru channel, create hole substitution by reducing electron density and modify the electronic correlation effect which would have significant ramification on the magnetic and electronic properties of material. Therefore, one can expect an increase in $U$ and decrease in effective density of states at Fermi level $N(\epsilon_F$) in SrRuO$_3$ with substitution concentration. The photo-emission spectroscopy studies, indeed, have indicated a moderate $U$ in SrRuO$_3$ which increases while $N(\epsilon_F$) depletes with progressive substitution of Ti.\cite{kim-p} The band structure calculation has additionally demonstrated that the on-site $U$ increases with Ti substitution in SrRuO$_3$ which would likely to induce metal-insulator transition in present SrRu$_{1-x}$Ti$_x$O$_3$ series.\cite{abbate} The electrical transport measurements in films of SrRu$_{1-x}$Ti$_x$O$_3$ show a metal-insulator transition at $x$ $\sim$ 0.5 and the system evolves through diverse electronic phases with variation of Ti concentration.\cite{kim-mit,lin} Recently, we have studied the evolution of magnetic behavior in SrRu$_{1-x}$Ti$_x$O$_3$ where the non-change of $T_c$ across the series has been explained with an opposite change of $U$ and $N(\epsilon_F$) with $x$ within the model of itinerant FM.\cite{gupta1,gupta2}

Here, we have focused on evolution of transport behavior by Ti substitution. Previous studies have shown that Ti substitution induces a metal-insulator transition around $x$ = 0.5 in SrRu$_{1-x}$Ti$_x$O$_3$, however, the detail study of electrical transport behavior in presence of magnetic field and the study related to evolution of electronic correlation is lacking. \cite{kim-mit,lin} Our study shows a metal-insulator transition is induced for $x$ $\geq$ 0.4. While the charge transport in PM-metallic (up to $x$ $\leq$ 0.4) and in insulating state ($x$ $>$ 0.5) follows a modified Mott's variable range hopping (VRH) model, the FM-metallic state follows a $T^2$ and $T^{3/2}$ dependence and its crossover with temperature. The electronic coefficient of specific heat decreases but the Kadowaki-Woods ratio increases over the series indicating an increasing of electronic correlation effect with progressive substitution of Ti.

\section {Experimental Details}
The series of polycrystalline samples SrRu$_{1-x}$Ti$_x$O$_3$ with $x$ = 0.0, 0.1, 0.2, 0.3, 0.4, 0.5 and 0.7 have been prepared using solid-state reaction method. The details of samples preparation and characterization of ingredients and temperature used have been discussed elsewhere. \cite{gupta1,gupta2} Temperature and field dependent magnetic measurements have been done using SQUID. Further, the dc electrical resistivity $\rho(T)$ and magnetoresistance (MR) data are collected using a home-built setup attached with Oxford superconducting magnet by following the four probe technique, in the temperature. The low temperature specific heat $C_p(T)$ measurements have been done with a home-built setup following semi adiabatic method. 

\begin{figure}
\centering
		\includegraphics[width=8.5cm]{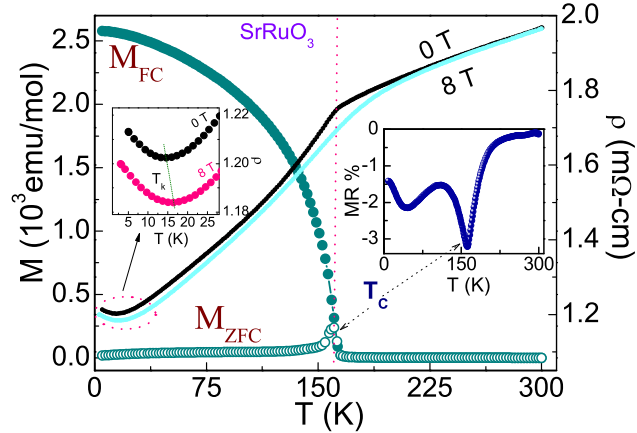}
\caption{The left axis shows temperature dependent dc magnetization data of SrRuO$_3$ measured in 100 Oe following zero field cooled (ZFC) and field cooled (FC) protocol. Data show a large bifurcation between FC and ZFC magnetization below the ferromagnetic ordering temperature $T_c$. Right axis shows the temperature dependent electrical resistivity measured in 0  and 8 T magnetic fields. Vertical dotted line represents $T_c$ of SrRuO$_3$. The left inset presents an expanded view of $\rho(T)$ at low temperature showing the minimum in $\rho(T)$ and its evolution with magnetic field. The right inset shows calculated magnetoresistance (Eq. 1) at 8 T field for SrRuO$_3$ as a function of temperature.}
	\label{fig:Fig1}
\end{figure}  

\section{Results and Discussions}
\subsection{Electrical Resistivity study}
The temperature dependent magnetic and electrical transport data of SrRuO$_3$ are shown in Fig. 1. The left axis of Fig. 1 shows the dc magnetization data of SrRuO$_3$ in temperature range between 5 K to 300 K where the data have been collected following field cooled (FC) and zero field cooled (ZFC) protocol in 100 Oe magnetic fields. The $M(T)$ data show a ferromagnetic (FM) to paramagnetic (PM) transition at temperature $T_c$ $\sim$ 163 K which is second-order phase transition in nature. The $T_c$ is indicated by a vertical dotted line and arrow. The wide bifurcation between $M_{ZFC}$ and $M_{FC}$ below $T_c$ indicates a large anisotropy associated with SrRuO$_3$.\cite{palai,kanbayasi} The critical analysis of this PM-FM transition has shown a mean-field like magnetic interaction in SrRuO$_3$.\cite{gupta1,gupta2,fuchs,cheng,kim-mf,gupta3} The right axis of Fig. 1 shows temperature dependent electrical resistivity $\rho(T)$ of SrRuO$_3$, measured in 0 and 8 T magnetic field. The $\rho(T)$ of SrRuO$_3$ shows metallic behavior throughout the temperature range. Interestingly, the $\rho(T)$ shows a slope change around $T_c$ which implies a close correlation between magnetic and transport behavior in SrRuO$_3$. Above $T_c$ in PM region, the $\rho(T)$ increases linearly without any saturation till temperature as high as 1000 K.\cite{allen} Even, $\rho(T)$ crosses the Ioffe-Regel limit for conductivity, which is an indication for bad metallic nature in SrRuO$_3$.\cite{gurvitch,klein-bm,emery} Below $T_c$ in FM region, the $\rho(T)$ decreases sharply which is attributed to spin disorder effect. Further, a minimum in $\rho(T)$ (indicated by a dotted circle in Fig. 1) has been observed around 15 K ($T_K$), where $\rho(T)$ increases as the temperature is lowered (discussed later). The $\rho(T)$, however, does not show any major change in magnitude from 5 K (1.21 m$\Omega$-cm) to 300 K (1.98 mΩ$\Omega$-cm) at zero magnetic field. The residual resistivity ratio (RRR), $\rho$(300 K)/$\rho$(5 K), for SrRuO$_3$ is obtained to be $\sim$ 1.62 which matches with other study. By applying of magnetic field (8 T), the $\rho(T)$ of SrRuO$_3$ decreases that gives a negative magnetoresistance (MR). The kink in $\rho(T)$ around $T_c$ is suppressed in presence of magnetic field which is due to broadening of magnetic transition. The MR has been calculated using following relation, 

\begin{eqnarray}
	MR\% = \frac {\Delta \rho}{\rho(0)} \times 100 = \left[\frac{\rho (H)-\rho (0)}{\rho (0)}\right] \times 100
\end{eqnarray}

where $\rho(H)$ is the resistivity in magnetic field and $\rho(0)$ is the resistivity recorded in zero magnetic field. The inset of Fig. 1 shows the calculated negative MR with pronounced dips around $T_c$ and at low temperature $\sim$ 50 K which is also matches with reported studies. \cite{moran,sow}

The $\rho(T)$ data for SrRu$_{1-x}$Ti$_x$O$_3$ series are shown in Fig. 2a and Fig. 2b, respectively. It is evident in Fig. 2a that with an increase of Ti substitution, the $\rho(T)$ increases but shows an identical metallic behavior till $x$ $\sim$ 0.3. Further, Fig. 2a shows a temperature driven metal to insulator transition (MIT) around 74 K ($T_{MIT}$) for $x$ = 0.4 sample. The insulating behavior in SrRu$_{1-x}$Ti$_x$O$_3$ series increases as the Ti substitution concentration increases for $x$ = 0.5 and 0.7, shown in Fig. 2b.\cite{kim-p,abbate,kim-mit,maiti} It may be noted that for samples with higher $x$ values, the increased magnitude of resistivity probably arises due to less number of Ru-O-Ru conducting paths as Ti$^{4+}$ substitution acts for site dilution for Ru$^{4+}$. The magnitude of $\rho(T)$ at low temperature (5 K) increases from 1.21 m$\Omega$-cm and 62.89 m$\Omega$-cm for $x$ = 0.7 sample indicating a substantial change towards the insulating behavior. The $\rho(T)$ does not saturate at high temperature for metallic samples ($x$ = 0.0, 0.1, 0.2, 0.3 and 0.4) and the derivative d$\rho$/dT (not shown here) remains finite down to the lowest temperatures. The ground state of $\rho(T)$ switches from FM-metallic to FM-insulating with Ti substitution around $x$ = 0.4.

\begin{figure}
	\centering
		\includegraphics[width=8.5cm]{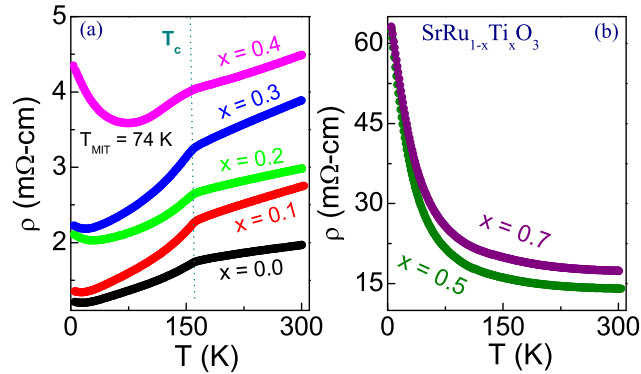}
	\caption{The temperature dependent electrical resistivity $\rho(T)$ have been shown for SrRu$_{1-x}$Ti$_x$O$_3$ series with (a) $x$ = 0.0, 0.1, 0.2, 0.3, 0.4 and (b) $x$ = 0.5 and 0.7 in temperature range from 5 K to 300 K.}
	\label{fig:Fig2}
\end{figure} 

The metal-insulator transition has been a subject of intense research in condensed matter physics for long time. There are many possible reasons for metal-insulator transition such as, disorder, electron correlation, hole substitution, electron-phonon coupling, etc. The sources of disorder in real materials are many that include lattice impurities, crystal defects, chemical substitution and inhomogeneity, etc. Disorder usually provides random potential in the system that leads to localization of electronic wave-function, hence Anderson like insulating phenomena is realized. The strong electron correlation effect, on the other hand, induces Mott like insulating state in a material even with partially filled band(s). Similarly, hole substitution amounts to depletion of density of state across Fermi level which results in a metal-insulator transition behavior. The effect of both disorder and electron correlation on electronic conduction mechanism have been studied quite intensively for last few decades. In present SrRu$_{1-x}$Ti$_x$O$_3$ series, Ti$^{4+}$ replaces Ru$^{4+}$ which has 3$d^0$ and 4$d^4$ electronic structure, respectively. The immediate effects of present substitution are it introduces hole substitution, increases electron correlation effect and induces site disorder creating local potential where all are in favor of transition towards an insulating phase. The SrRuO$_3$ is commonly believed to a correlated metal sitting on edge of Mott transition while the other end member i.e., SrTiO$_3$ is a band insulator with an energy gap around 3.7 eV. \cite{allen,groot} Previous study on series of SrRu$_{1-x}$Ti$_x$O$_3$ films has shown metal-insulator transition occurs around $x$ = 0.5, and various electronic states such as, correlated metal ($x$ = 0), disordered metal ($x$ = 0.3), Anderson insulator ($x$ = 0.5), Coulomb gap insulator ($x$ = 0.6), disordered correlation insulator ($x$ = 0.8) and band insulator ($x$ = 1.0) are observed across the series.\cite{kim-mit} Our study shows a metal-insulator transition in present SrRu$_{1-x}$Ti$_x$O$_3$ series at $x$ around 0.4 which is consistent with previous study, even considering the fact that films are accompanied by substrate strain and electronic confinement. The photo-emission spectroscopy study has, however, shown an opening of soft gap at Fermi level around $x$ $\sim$ 0.5 and a hard gap at higher values of $x$ where the metal-insulator transition is ascribed to an increasing amount of local disorder and electron correlation in the system. Nonetheless, the present series offer an ideal background to study the combined effect of disorder and electronic correlation on charge conduction on oxide system. Interestingly, it can be further mentioned that the $x$ value for metal-insulator transition in present series is close to percolation threshold ($\sim$ 30\%) for 3-dimensional simple cubic lattice.\cite{sykes}

\setlength{\tabcolsep}{4pt}
\begin{table}
\caption{\label{tab:table I} The coefficients ($\rho_{0M}$ and $T_M$) of modified Mott's VRH model (Eq. 2) are shown with variation of Ti concentration in SrRu$_{1-x}$Ti$_x$O$_3$ series.}

\begin{indented}
\item\begin{tabular}{c|c|c|c}
\hline
Ti(x)  &Field &$\rho_{0M}$ (m$\Omega$-cm K$^{-1/2}$) &$T_M (K)$\\ 
\hline
0.0 &0 T &0.037 &466\\
  &8 T &0.038 &439\\
\hline
0.1 &0 T &0.071 &122\\
  &8 T &0.077 &078\\
\hline
0.2 &0 T &0.052 &583\\
  &8 T &0.053 &552\\
\hline
0.3 &0 T &0.097 &145\\
  &8 T &0.106 &093\\
\hline
0.4 &0 T &0.076 &646\\
  &8 T &0.086 &426\\
\hline
0.5 &0 T &0.063 &12642\\
  &8 T &0.064 &12458\\
\hline
0.7 &0 T &0.070 &14561\\
  &8 T &0.074 &13449\\
\hline
\end{tabular}
\end{indented}
\end{table}

\subsection{Electrical Transport in PM State}
In this section, we discuss about the charge transport mechanism in high temperature PM state for SrRu$_{1-x}$Ti$_x$O$_3$ series. Fig. 2a shows $\rho(T)$ in PM state ($T$ $>$ $T_c$) continuously increases almost linearly till $x$ = 0.4. On the other hand, Fig. 2b shows $\rho(T)$ continuously decreases showing an insulating behavior. For insulating disordered materials, electronic transport usually occurs due to thermally activated hopping of charge carriers among localized states around Fermi level, therefore the detail nature of density of states across Fermi level plays an important role. In most of the cases charge conduction follows Mott's variable range hopping (VRH) model which considers hopping of trapped charge carriers in disorder materials and assumes a nearly constant density of states.\cite{mott} The disorder is, however, another parameter which has large influence on the charge conduction mechanism. To take an account of the role of disorders in materials, the originally proposed Mott's VRH model has further been modified by Greaves including a $\sqrt{T}$ term.\cite{greaves} According to Greaves prescription, the modified Mott's VRH model is given by 

\begin{eqnarray}
	\rho(T) = \rho_{0M} \sqrt{T} exp[(T_M/T)^{1/4}]
\end{eqnarray}

where $\rho_{0M}$ is the pre-exponential constant due to the electron-phonon interaction and $T_M$ is the Mott's characteristic temperature that depends on density of states. The $T_M$ basically measures the degree of disorder in system and its relation with $\rho_{0M}$ at Fermi level is given by through N($\epsilon_F$),

\begin{eqnarray}
	\rho_{0M} = \frac{1}{3\nu_{ph} e^2} \left[\frac{8\pi \alpha k_B} {N(\epsilon_F)}\right]^{1/2}
\end{eqnarray} 

\begin{eqnarray}
T_M = 19.45 \left[\frac{\alpha^3}{N(\epsilon_F)k_B}\right]
\end{eqnarray}

where $\nu_{ph}$ is the optical phonon frequency, $k_B$ is the Boltzmann constant, $e$ is the electronic charge and $\alpha$ (= 1/$\xi$) is the inverse localization length of the localized states. The $\rho(T)$ data both in high temperature paramagnetic-metallic and in insulating state have been fitted well with Eq. 2 in Figs. 3a and 3b, respectively. Here, we mention that tried to fit our $\rho(T)$ data with all the available thermally activated hopping models but we find Eq. 2 is more suitable in terms of fitting range as well as fitting indicator. The same model (Eq. 2) has been used by Kim \textit{et al.}\cite{kim-mit} to fit the $\rho(T)$ in insulating state of SrRu$_{1-x}$Ti$_x$O$_3$ ($x$ = 0.5), terming it as an Anderson insulator highlighting the role of disorder on charge conduction mechanism. Here, it can be noted that single Eq. 2 can be used to explain the charge transport in PM state both for metallic ($x$ $<$ 0.4) as well as insulating ($x$ $>$ 0.4) samples. The validity of Eq. 2 is extended in FM state for insulating samples. It is evident in Fig. 3b that Eq. 2 can be fitted in whole temperature range for samples with high Ti concentration. The obtained fitting parameters such as, $\rho_{0M}$ and $T_M$ are given in Table 1 for the present series. In PM-metallic state the $T_M$ changes nonmonotonically showing an oscillatory change till $x$ = 0.4, but its value increases significantly by order of two in case of insulating materials ($x$ = 0.5 and 0.7). While the electronic density of state $N(\epsilon_F)$ ($\propto$ electronic coefficient to specific heat $\gamma$, Eq. 11) decreases continuously with $x$, the nonmonotonic changes of $T_M$ is surprising,  which probably arises due to a nonmonotonic change of $\alpha$ in Eq. 4. The substantial increase of $T_M$, however, in insulating materials is due to significant depletion of density of states in higher substituted samples.  

\begin{figure}
	\centering
		\includegraphics[width=8.5cm]{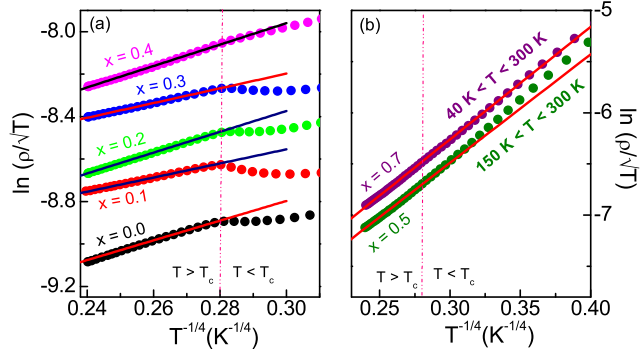}
	\caption{Temperature dependent resistivity data are plotted following modified form of Mott's VRH model (Eq. 2) for SrRu$_{1-x}$Ti$_x$O$_3$ series and solid lines are due to straight line fitting in high temperature range of with compositions (a) $x$ = 0.0, 0.1, 0.2, 0.3, 0.4 and (b) 0.5 and 0.7.}
	\label{fig:Fig3}
\end{figure}

\setlength{\tabcolsep}{0.5pt}
\begin{table*} 
\caption{\label{tab:table II} The coefficients $A$, $B$ and $C$ in Eq. 5, Eq. 7 and Eq. 8, respectively are shown along with their effective temperature range of fitting for SrRu$_{1-x}$Ti$_x$O$_3$ series.}

\begin{indented}
\item\begin{tabular}{ccccccc}
\hline
Field \ &Property &Parameter &$x$ = 0.0 &0.1 &0.2 &0.3\\
\hline
0 T &$T^2$ &$\rho_{0A}$ (m$\Omega$-cm) &1.25 &1.37 &1.93 &2.18\\
 & &$A$ (m$\Omega$-cm K$^{-2}$) $\times$ 10$^{-5}$ &1.89 &3.48 &2.81 &4.25\\
 & &T Range (K) &67 - 163 \ &119 - 163 \ &117 - 163 \ &91 - 163\\
\hline
0 T &$T^{3/2}$ &$\rho_{0B}$ (m$\Omega$-cm) &1.18 &1.31 &1.90 &2.11\\
& &$B$ (m$\Omega$-cm K$^{-3/2}$) $\times$ 10$^{-4}$ &2.98 &4.33 &3.30 &4.96\\
 & &T Range (K) &20 - 70 \ &38 - 131 \ &85 - 123 \ &41 - 106\\
\hline
8 T &$T^{3/2}$ &$\rho_{0B}$ (m$\Omega$-cm) &1.17 &1.25 &1.86 &1.98\\
 & &$B$ (m$\Omega$-cm K$^{-3/2}$) $\times$ 10$^{-4}$ &2.58 &4.44 &3.42 &5.60\\
 & &T Range (K) &27 - 163 \ &47 - 163 \ &87 - 163 \ &83 - 163\\
\hline
0 T &Kondo &$\rho_{0K}$ (m$\Omega$-cm) &1.23 &1.38 &2.17 &2.27\\
 & &$C$ (m$\Omega$-cm) $\times$ 10$^{-2}$ &1.19(2) &1.52(9) &3.93(28) &2.96(11)\\
\hline
8 T &Kondo &$\rho_{0K}$ (m$\Omega$-cm) &1.21 &1.34 &2.12 &2.20\\
 & &$C$ (m$\Omega$-cm) $\times$ 10$^{-2}$ &1.12(5) &1.50(8) &3.56(10) &2.83(16)\\
\hline
\end{tabular}
\end{indented}
\end{table*}

\subsection{Electrical Transport in FM state}
For SrRuO$_3$, the $\rho(T)$ below $T_c$ in FM state also shows metallic behavior, however, a distinct slope change is evident across $T_c$ (Fig. 1). This implies that magnetic ordering has profound influence on the charge conduction mechanism which reduces spin disorder scattering. The charge conduction below $T_c$ has been observed to follow the functional form, 

\begin{eqnarray}
  \rho(T) = \rho_{0A} + AT^2
\end{eqnarray}

where $\rho_{0A}$ is the residual resistivity due to impurity scattering and $A$ is the coefficient which signifies the scattering rate. The $T^2$ dependence of $\rho$ is an indication of dominant electron-electron interactions forming a Fermi liquid (FL) state which gives a different temperature dependence of $\rho(T)$, compared to simple metals.\cite{landau}

\begin{figure}
	\centering
		\includegraphics[width=8cm]{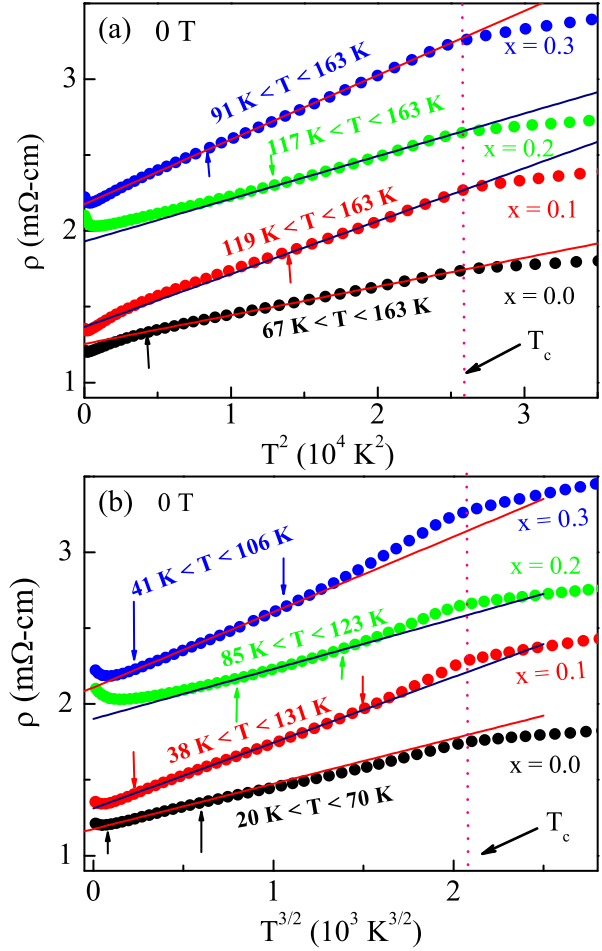}
	\caption{(a) Temperature dependent resistivity data have been plotted and fitted with Eq. 5 below $T_c$ for SrRu$_{1-x}$Ti$_x$O$_3$ series with $x$ = 0.0, 0.1, 0.2, and 0.3. (b) The same have been shown and the straight lines are due to fitting with Eq. 7 at low temperature region. The dotted lines represent $T_c$.}
	\label{fig:Fig4}
\end{figure} 

We have plotted the resistivity of metallic samples ($x$ $\leq$ 0.3) as a function of $T^2$, as shown in Fig. 4a. It is evident in Fig. 4a that $\rho(T)$ of SrRuO$_3$ follows a quadratic temperature dependence below $T_c$. However, this $T^2$ dependence deviates at low temperature which is marked by vertical arrows in Fig. 4a. For Ti substituted materials ($x$ = 0.1, 0.2 and 0.3), we also observe similar $T^2$ dependence below $T_c$, however, the range of fitting modifies with Ti substitution. The temperature range for $T^2$ dependence as well as the values of $\rho_{0A}$ and $A$ (Eq. 5) have been shown in Table 2. For SrRuO$_3$, $\rho(T)$ follows a $T^2$ dependence down to $\sim$ 67 K. In substituted materials the fitting range decreases while keeping upper temperature $T_c$ fixed. This suggests that breakdown of $T^2$ dependence is triggered by other phenomenon, active at low temperatures. Note, that parameter $\rho_{0A}$ shows a comparatively higher value than that for single crystals or thin films but the value of $A$ closely matches with those values.\cite{allen,wang}

In parallel to Fermi liquid behavior, the $T^2$ dependence of $\rho(T)$ has also been discussed to originate at low temperatures due to prominent electron-magnon ($e$-$m$) scattering with following temperature dependence,\cite{wang}

\begin{eqnarray}
  \rho_{e-m} = \frac{3\pi^5S}{16e^2}\left(\frac{\mu k_B N J_0}{m_e E_F^2}\right)^2\frac{h}{2\pi k_F}T^2
\end{eqnarray}

where $\mu$ is an effective mass of magnon, $NJ_0$ is the spin-orbit coupling constant, $E_F$ is the Fermi energy and the $k_F$ is the Fermi wave vector. The $T^2$ coefficient ($A_{e-m}$) in Eq. 6 has been calculated to be around 4.15 $\times$ 10$^{-7}$ m$\Omega$-cm K$^{-2}$ for SrRuO$_3$ taking an appropriate value of $E_F$, Fermi velocity $V_F$ and $NJ_0$.\cite{wang} This calculated $A_{e-m}$ shows a low value than our obtained $A$ in Table 2. Here, it can be noted that this $T^2$ dependence of $\rho(T)$ is evident at low temperatures (favorably below 40 K) in other studies of SrRuO$_3$,\cite{allen,wang} but we observe this dependence at higher temperature below $T_c$. Nonetheless, a sudden slope change in $\rho(T)$ across $T_c$ a significant fractional value of $A_{e-m}$ imply that this electron-magnon scattering has contribution to $T^2$ dependence of $\rho(T)$ along with electron-electron scattering related to FL behavior. Given that the electron correlation is the key factor in FL behavior, modification of $U$ as well as electron density with Ti$^{4+}$ (3$d^0$) substitution will have a significant influence on electron transport behavior, hence on the observed $T^2$ dependence. For instance, coefficient $A$, which signifies the quasiparticle scattering rate, is expected to increase with the decrease of electronic density. The Table 2 shows an increasing $A$ with $x$ which can be explained with the depletion of charge carriers as Ti$^{4+}$ (3$d^0$) is introduced in the system.  

\begin{figure}
	\centering
		\includegraphics[width=8.5cm]{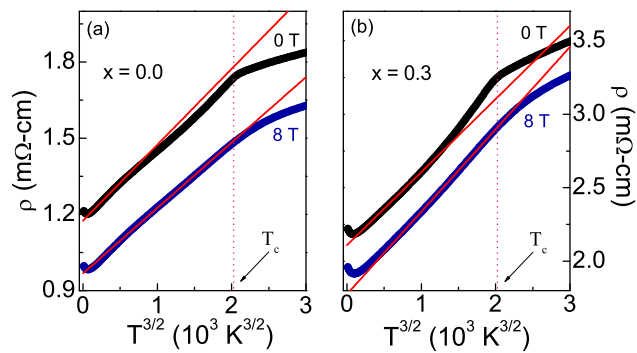}
		\caption{The resistivity vs temperature data in 0 and 8 T field are shown for representative (a) $x$ = 0.0 and (b) $x$ = 0.3 material of SrRu$_{1-x}$Ti$_x$O$_3$ series. The $\rho(T)$ of 8 T has been shifted vertically by 0.2 for clarity. The straight lines are due to fitting with Eq. 7.}
	\label{fig:Fig5}
\end{figure}  

Following a deviation of $T^2$ dependence at low temperature, we have analyzed the $\rho(T)$ with following temperature dependence form,

\begin{eqnarray}
  \rho(T) = \rho_{0B} + B T^{3/2}
\end{eqnarray}

where $\rho_{0B}$ is the temperature independent residual resistivity and $B$ is the coefficient. The $T^{3/2}$ dependence in $\rho(T)$ is mostly attributed to non-Fermi liquid (NFL) behavior which naturally arises with the breakdown of FL behavior. However, other theoretical models such as, antiferromagnetic spin fluctuations\cite{moriya} or incoherent electron-magnon scattering\cite{mills} have also explained this unconventional $T^{3/2}$ dependence of resistivity. While the former model can be excluded for this present FM SrRuO$_3$, the later one arises from a nonconserving-wave-vector (incoherent part) scattering of electron-magnon showing a prominent effect at low temperature below the characteristic temperature $T_m$ (= $\epsilon_m$/$k_B$, where $\epsilon_m$ is magnon energy). The coherent part of electron-magnon scattering, on the other hand, is not significant at low temperature because there are no available long-wave-vector thermal magnons to participate in scattering process with electrons. This model\cite{mills} further suggests a strengthening of $T^{3/2}$ dependence with disorder. Fig. 4b presents the plotting of $\rho(T)$ vs $T^{3/2}$ showing a linear dependence at low temperature, below the temperature range where a $T^2$ dependence has been observed (Fig. 4a). Table 2 shows that the $T^{3/2}$ dependence is valid in temperature range of 20 to 70 K for SrRuO$_3$, but in Ti substituted samples this range extends to higher temperatures.

The appearance of NFL behavior usually occurs in the strange metals such as, heavy fermion materials close to quantum critical point, copper oxide based superconductors, etc.\cite{stewart,lee,gegenwart,boebinger} The prominent example of FL to NFL crossover with substitution is system like Sr$_{1-x}$Ca$_x$RuO$_3$ which shows a quantum-phase-transition across $x$ = 0.7.\cite{cao,fuchs} The NFL behavior has previously been reported in SrRuO$_3$ through optical conductivity/reflectivity measurements by Kostic \textit{et al.},\cite{kostic} and that has been further theoretically explained by Laad \textit{et al.}\cite{laad} from its structural organization. SrRuO$_3$ has complex transport behavior where both FL and NFL behavior have been observed by several groups but in most cases FL behavior ($T^2$ dependence) has been observed at low temperatures, in contrast to present study.\cite{klein-bm,klein-rrr,wang,mackenzie,ahn,dodge1,dodge2,klein-nfl} The $\rho(T)$ in SrRuO$_3$ shows a continuous linear increase (till at least 1000 K) crossing the Ioffe-Regel limit which is considered as metallic conductivity limit based on system's lattice parameters. The FL to NFL crossover in these studies has been attributed to this bad metallic character of SrRuO$_3$.

The crossover from $T^2$ to $T^{3/2}$ dependence at low temperature in present series is quite noteworthy. We believe that this breakdown of $T^2$ dependence or appearance of $T^{3/2}$ is likely to be caused by an incoherent part of electron-magnetic scattering.\cite{mills} An estimation of $T_m$ $\sim$ 70 K for SrRuO$_3$ by Mazin and Singh\cite{mazin}matches with the upper temperature limit (70 K) for $T^{3/2}$ dependence in present SrRuO$_3$ system (Table 2). Following this model, $T^{3/2}$ dependence should extend to higher temperature in presence of magnetic field as $\epsilon_m$ is expected to increase with the rigidity of spin ordering against thermal fluctuations.\cite{mills} As expected, Fig. 5 shows $T^{3/2}$ dependence is extended up to $T_c$ in 8 Tesla field for representative $x$ = 0 and 0.3 samples. The obtained fitted parameters $\rho_{0B}$ and $B$ are shown in Table 2. While the coefficient $B$ shows a slight increase with Ti, it does not change appreciably with magnetic field. However, our $B$ parameter is roughly one order higher than the case where $T^{3/2}$ dependence has been observed at high temperature ($>$ 50 K) ascribing to NFL behavior.\cite{wang} Nonetheless, the observation of both $T^2$ and $T^{3/2}$ dependence and its crossover with temperature in the present series is quite noteworthy. 

\begin{figure}
	\centering
		\includegraphics[width=8cm]{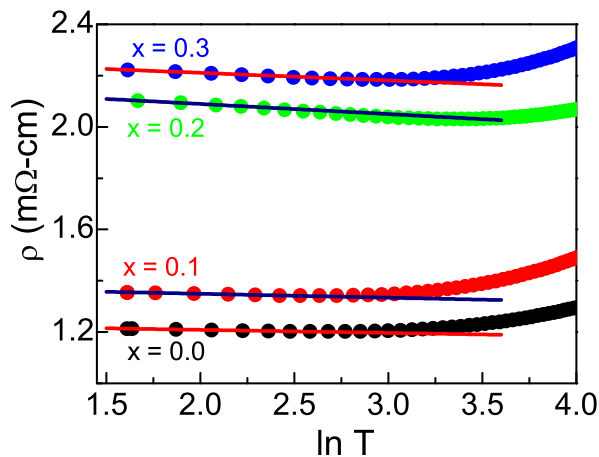}
	\caption{Temperature dependent resistivity are plotted in semi-log scale at low temperature for SrRu$_{1-x}$Ti$_x$O$_3$ series with $x$ = 0.0, 0.1, 0.2, and 0.3. The straight lines are due to fitting following Kondo behavior (Eq. 8).}
	\label{fig:Fig6}
\end{figure}
											
\subsection{Kondo Behavior}
It is seen in Fig. 1 that $\rho(T)$ for SrRuO$_3$ shows a small upturn at low temperature below $\sim$ 14.6 K ($T_K$). This upturn or minimum in $\rho(T)$, where resistivity at low temperature increases with decreasing temperature, arises due to various reasons such as, Kondo effect \cite{kondo,mydosh}, weak localization effect \cite{tvr,herranz}, intergrain transport \cite{roy}, etc. The intergrain transport assumes charge transport across the grain boundaries where in zero magnetic field the spins in neighboring grains have non-parallel alignment which restricts the charge carrier movement, giving an increasing resistivity at low temperature. With increasing temperature, the thermal energy helps for spin reorientation that results in a minimum in $\rho(T)$. This minimum in $\rho(T)$ due to intergrain transport mechanism is highly sensitive to the applied magnetic field which lowers the minimum toward lower temperature or even vanishes the minimum.\cite{roy} Weak localization effect, on the other hand, mostly arises in disordered materials due to interference of electronic wavefunctions when they are coherently backscattered by randomly distributed scattering centers. In Weak localization effect, the conductivity ($\sigma$) follows a $T^{1/2}$ dependence. In present SrRuO$_3$, we find $\sigma$ $\propto$ $T^{1/2}$ dependence is followed up to around 8 K (not shown) which is much lower than the $T_K$. The left inset of Figure 1 shows the minimum in $\rho(T)$ or $T_K$ does not shift significantly with magnetic field. For instance, in magnetic field as high as 8 T the $T_K$ shifts to higher temperature only by $\sim$ 1.5 K. This mostly insensitive behavior of $T_K$ and the nature of shifting of $T_K$ in magnetic field implies this minimum in $\rho(K)$ is unlikely due to intergrain transport or weak localization effect. The Kondo effect arises due to scattering of conduction electrons with the magnetic impurities resulting in a continuous increase in resistivity at low temperature.\cite{kondo,mydosh} In case of Kondo behavior, resistivity follows logarithmic of temperature dependence with following functional form,

\begin{eqnarray}
	\rho(T) = \rho_{0K} - C \ln T
\end{eqnarray} 

where $\rho_{0K}$ is the Kondo residual resistivity (i.e., sample impurity) and $C$ is the Kondo coefficient. Following Eq. 8, we have plotted $\rho(T)$ vs $\ln T$ in Fig. 6 where the linear behavior close to $T_K$ indicates Kondo-like behavior till Ti substitution concentration $x$ = 0.3. As we increase the Ti substitution level ($x$ $\geq$ 0.3), $\rho(T)$ shows an insulating behavior and the Kondo effect disappears which also suggests this minimum in $\rho(T)$ is unlikely due to intergrain transport mechanism. Recently, the possibility of Kondo behavior has been discussed in SrRuO$_3$ films.\cite{ghosh} The parameters ($\rho_{0K}$ and $C$) obtained from fitting of Eq. 8 are given in Table II in 0 and 8 T field for $x$ $\leq$ 0.3. As evident, both the parameters do not change appreciably and remain nearly constant within the limit of error bar. This field-independent nature of coefficient $C$ is further regarded as typical feature of Kondo behavior \cite{ghosh,roy,he}. Therefore, this low temperature minimum in $\rho(T)$ in present series ($x$ $\leq$ 0.3) is likely due to Kondo-like behavior.
       
\begin{figure}
	\centering
		\includegraphics[width=8.5cm]{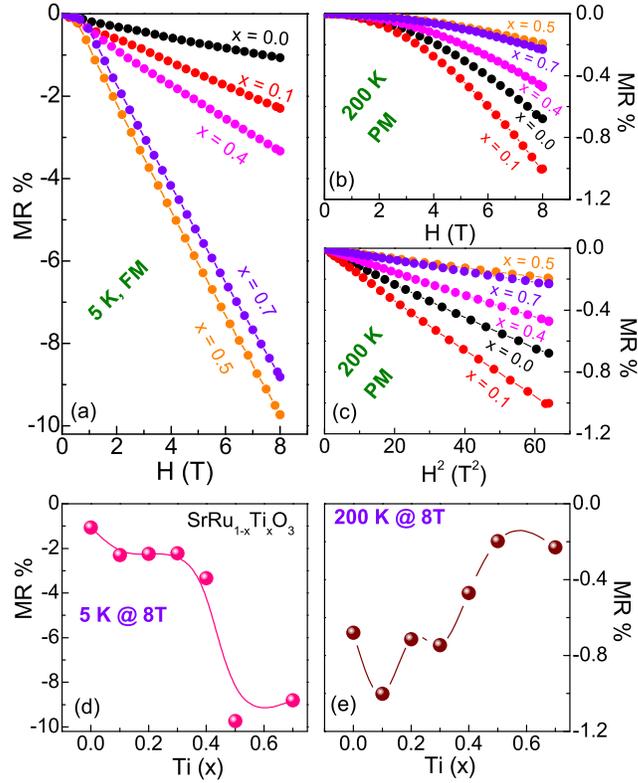}
	\caption{The magnetoresistance (Eq. 1) with variation of field are shown (a) at 5 K and (b) at 200 K for SrRu$_{1-x}$Ti$_x$O$_3$ series. (c) shows MR at 200 K with quadratic dependance of field. (d) and (e) show the value of MR at 8 T field of SrRu$_{1-x}$Ti$_x$O$_3$ series at 5 K and 200 K, respectively.}
	\label{fig:Fig7}
\end{figure}

\subsection{Magnetoresistance}
The electrical resistivity have further been measured in presence of magnetic field. The MR calculated for SrRuO$_3$ shows a negative value throughout the temperature range where two prominent dips around $T_c$ and 50 K are observed (Fig. 1).\cite{moran,sow} The dip in MR across $T_c$ is associated with broadening of phase transition where $\rho(T)$ shows a smooth change. Magnetic field dependent MR have been measured at different temperatures to understand the effect of magnetic state on charge transport. For instance, we have plotted the isothermal MR for present series at two selected temperatures i.e., 5 K and 200 K, which represents FM and PM regions, respectively in Figures 7a and 7b. A negative MR has been observed at both these temperatures. The MR varies linearly with the applied magnetic field in FM regime at 5 K (Fig. 7a) while in PM state, this variation is nonlinear for all the samples (Fig. 7b). However, a linear variation of MR in PM state at 200 K is observed with square of magnetic field in Fig. 7c. The quadratic field dependance of MR has been observed for both metallic as well as nonmetallic samples without spin ordering (PM state) at high temperatures. In PM state above $T_c$, MR $\propto$ $H^2$ basically implies a MR $\propto$ $M^2$ due to linear relationship $M$ = $\chi H$ where $M$ is the magnetic moment and $\chi$ is the magnetic susceptibility. In PM state, as the charge carriers are scattered by thermally fluctuating spins, this increasing moment with magnetic field reduces the spin fluctuations which in turn increases the charge conductivity, hence a negative MR is realized. In Fig. 7d, we have shown the MR value observed at highest magnetic field of 8 T for different Ti concentration in FM region at 5 K. For the metallic samples till $x$ = 0.3, the MR is not impressive and its value does not change much with $x$. With the introduction insulating phase ($x$ $\geq$ 0.4) however, MR increases even in same FM state, implying spin ordering with magnetic field promotes charge conduction. Interestingly, an opposite evolution of MR has been observed at high temperature PM state where MR value decreases with $x$, though its values are not significantly different. This can be explained due to site dilution effect. This signifies the role of magnetic ordering and magnetic field on the charge transport behavior in present SrRu$_{1-x}$Ti$_x$O$_3$ series.

\begin{figure}
	\centering
		\includegraphics[width=8.5cm]{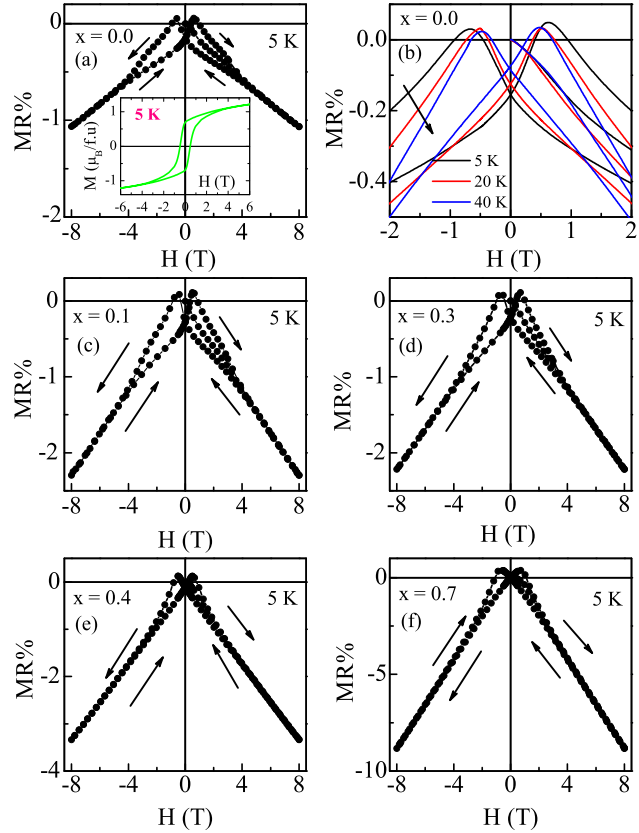}
	\caption{Magnetic field dependent MR data at 5 K are shown for (a) $x$ = 0.0, (c) $x$ = 0.1, (d) $x$ = 0.3, (e) $x$ = 0.4 and (f) $x$ = 0.7 of SrRu$_{1-x}$Ti$_x$O$_3$ series.  The arrows show the direction of field sweeping. The inset of (a) shows magnetic hysteresis $M(H)$ at 5 K. (b) shows MR data for SrRuO$_3$ at 5, 20 and 40 K in low field regime.}
	\label{fig:Fig8}
\end{figure}

Fig. 7e shows MR ratio at 8 T magnetic field for different Ti substitution concentration at 200 K in the paramagnetic region, as evaluated from Fig. 7b. It is evident from the data that in PM region MR does not have a significant change in magnitude as compared to FM regime. In PM region, a maximum MR around 1\% is observed for $x$ = 0.1 and the MR value decreases with increase in Ti substitution concentration to 0.2 \% for $x$ = 0.7. 

Further, the evolution of MR has been studied with sweeping of magnetic field from 0 to +8 to 0 T and then to -8 to 0 to +8 T at 5 K, indicated by arrows in Fig. 8. For SrRuO$_3$, MR is observed to be negative and it shows a reasonable hysteresis below $\sim$ 4 T. It also shows a remnant MR at zero field after returning from higher field. Interestingly, when field is reversed toward negative direction, the negative MR continue to decrease and the MR shows a small positive value and then with further increase of field, MR becomes again negative. The field where this positive MR arises closely matches with the coercive field $H_c$ of magnetic hysteresis loop. This shows positive MR arises due to weakening of moment at low temperature. Similar hysteresis and remnant value in MR has also been observed in negative field cycle (Fig. 8a). This behavior of MR appears to be associated with the nature of magnetism in the material. For instance, the inset of Fig. 8a shows the magnetic hysteresis loop $M(H)$ of SrRuO$_3$ at 5 K which similarly shows a wide hysteresis below $\sim$ 4 T as well as high remnant magnetization at zero magnetic field. This underlines that the spin ordering and charge transport behavior in SrRuO$_3$ is closely related. While returning from high magnetic field (8 T), system retains its induced moment which results in a higher negative MR and a remnant MR, this causes hysteresis in both $M(H)$ and MR. With increasing temperature, both the hysteresis as well as remnant MR decreases due to softening of spin ordering (Fig. 8b). We, however, have not found hysteresis in high temperature PM state (not shown). Further, with dilution of spin ordering through Ti substitution, we find both the hysteresis and remnant MR reduces, as seen in Figs. 8c, 8d, 8e and 8f for $x$ = 0.1, 0.3, 0.4 and 0.7, respectively. Here, it can be noted that with increasing temperature the similar decrease in hysteresis and remnant in MR have been observed for Ti substituted samples (not shown). This signifies the dominant role of magnetism on transport behavior in SrRuO$_3$.         
	
\begin{figure}
	\centering
		\includegraphics[width=8.5cm]{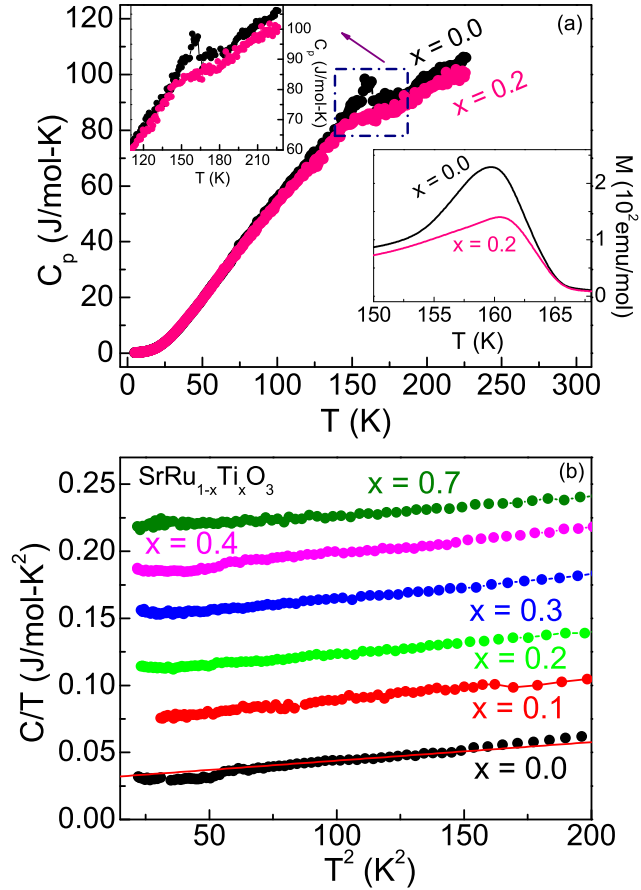}
	\caption{(a) Temperature dependent specific heat $C_p(T)$ are shown for SrRu$_{1-x}$Ti$_x$O$_3$ series with $x$ = 0.0 and 0.2. The upper and lower inset shows an expanded view of $C_p(T)$ and $M(T)$ data across $T_c$. (b) shows the $C_p(T)$ data plotted following Eq. 10 for SrRu$_{1-x}$Ti$_x$O$_3$ series with $x$ = 0.0, 0.1, 0.2, 0.3, 0.4 and 0.7. The line is due to straight line fitting. For clarity, the $C_p(T)$/T data have been shifted vertically by amount 0.02 for $x$ = 0.1, 0.2, 0.3, 0.4, and 0.7 samples.}
	\label{fig:Fig9}
\end{figure}
  
\subsection{Specific Heat}
The specific heat ($C_P$) has mainly been measured to understand an evolution of charge transport behavior and electron correlation effect in present series. In Fig. 9a, we have presented the temperature dependent specific heat for two representative samples i.e., $x$ = 0 and 0.2. It is clear in Fig. 9 that at low temperature ($<$ 20 K) $C_p(T)$ does not vary much, but above this temperature the $C_p(T)$ increases almost linearly. For $x$ = 0 parent compound, $C_p(T)$ exhibits a pronounced jump around $T_c$ which is shown in a magnified view in upper inset of Fig. 9a. The $C_p(T)$ for $x$ = 0.2 sample shows similar behavior at low temperature but the jump across $T_c$ is much softened. This implies that with Ti substitution, magnetic transition has been broadened, as also evidenced with a dip in MR across $T_c$ (see Fig. 1). Similar signature of broadening of transition has also been observed in temperature dependent magnetization data which are shown in lower inset of Fig. 9a. Here, it can be mentioned that similar behavior of $C_p(T)$ data has been observed in other materials in present SrRu$_{1-x}$Ti$_x$O$_3$ series. The specific heat has mainly two contributions namely, electronic and lattice contribution which has given below,       

\begin{eqnarray}
	C_p = \gamma T + \beta T^3    
\end{eqnarray} 

where $\gamma$ and $\beta$ are the electronic ($C_e$) and lattice ($C_l$) coefficients of specific heat, respectively. To identify the individual contribution of electronic and lattice part, the $C_p(T)$ data of Eq. 9 have been plotted as following,

\begin{eqnarray}
	C_p/T = \gamma + \beta T^2    
\end{eqnarray} 

The straight line fitting of $C_p(T)$ data using Eq. 10 gives $\gamma$ = 30 mJ mol$^{-1}$ K$^{-2}$ and $\beta$ = 0.138 mJ mol$^{-1}$ K$^{-4}$, respectively for SrRuO$_3$, where these values closely match with the reported values for this material.\cite{allen,cao} This high value of $\gamma$ suggests a considerable electronic correlation in SrRuO$_3$. The electronic density of states $N(\epsilon_F)$ has been calculated directly from $\gamma$ as,

\begin{eqnarray}
	\gamma = \frac{\pi^2 k_B^2 N_a N(\epsilon_F)} {3} 
\end{eqnarray} 

that gives $N(\epsilon_F)$ = 174 states/Ry f.u. for SrRuO$_3$ (for both spin directions) with $N_a$ is the Avogadro number. Similarly, the Debye temperature ($\Theta_D$) has been calculated from $\beta$ using following relation 

\begin{eqnarray}
	\Theta_D = \left[\frac {12 \pi^4 n R} {5 \beta}\right]^{1/3}    
\end{eqnarray}  

where $R$ = 8.314 J K$^{-1}$ mol$^{-1}$ is the universal molar gas constant and $n$ = 5 is the number of atoms per formula unit of SrRuO$_3$. Using Eq. 12, we have calculated $\Theta_D$ $\sim$ 413 K for SrRuO$_3$. The estimated values of $\gamma$ and $\Theta_D$ have been shown in Fig. 10a for present series where $\gamma$ or equivalent $N(\epsilon_F)$ (Eq. 11) and $\Theta_D$ decreases with progressive substitution of Ti. This behavior of $\Theta_D$ is likely due to substitution of lighter element for Ru. The decrease in density of states is reflected in an increase of resistivity with $x$ in SrRu$_{1-x}$Ti$_x$O$_3$ series, as shown in Fig. 2a. Here it can be noted that our result is in agreement with photo-emission data which similarly shows a decrease of density of states in SrRu$_{1-x}$Ti$_x$O$_3$ with $x$. \cite{kim-p}
	
\begin{figure}
	\centering
		\includegraphics[width=8.5cm]{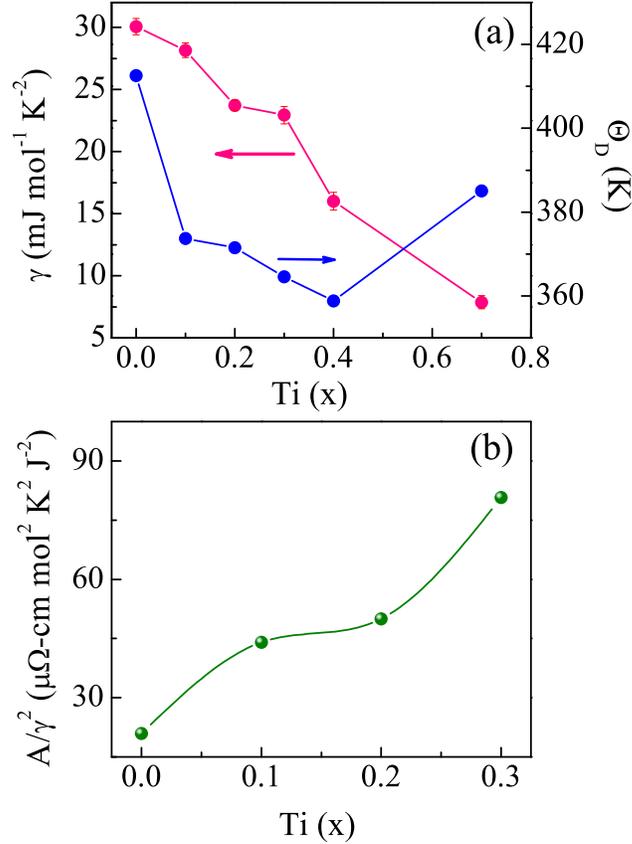}
	\caption{The electronic specific heat coefficient $\gamma$ (left axis) and the Debye temperature $\Theta_D$ (right axis) are shown for SrRu$_{1-x}$Ti$_x$O$_3$ series. (b) The Kadowaki-Woods ratio (see text) is shown for the metallic samples of present series.}
	\label{fig:Fig10}
\end{figure}

\subsection{Kadowaki-Woods Ratio}
The Kadowaki-Woods ratio (KWR), $A$/$\gamma^2$, is about comparing the coefficient of the quadratic term in $\rho(T)$ ($A$ in Eq. 5) with the coefficient of linear term in $C_p (T)$ ($\gamma$ in Eq. 9) at low temperature. According to FL model, the $A$ is squarely proportional to quasiparticle mass enhancement due to electron correlation effect. This KWR is believed to be suggestive of electron correlation strength in a material, and shows a constant or near constant value for a definite class of material.\cite{kadowaki} For instance, its value has been found to be 0.4 and 10 $\mu$$\Omega$-cm mol$^2$ K$^2$ J$^{-2}$ in case of transition metals and heavy fermions, respectively.\cite{kadowaki,rice,jacko,miyake}For SrRuO$_3$, using $A$ = 1.9 $\times$ 10$^{-5}$ m$\Omega$-cm K$^{-2}$ and $\gamma$ = 30 mJ mol$^{-1}$ K$^{-2}$, we calculate KWR to be around 21 $\mu$$\Omega$-cm mol$^2$ K$^2$ J$^{-2}$ that matches with other report.\cite{wang} This value of KWR for SrRuO$_3$ is almost double of that for heavy fermion systems but in the range of transition metal oxides.\cite{kadowaki} For example, La$_{1.7}$Sr$_{0.3}$CuO$_4$ shows KWR about 50 $\mu$$\Omega$-cm mol$^2$ K$^2$ J$^{-2}$, while a very high value around 500 $\mu$$\Omega$-cm mol$^2$ K$^2$ J$^{-2}$ has been observed in Na$_{0.7}$CoO$_2$. \cite{hussey-kw,li} With Ti concentration, KWR increases reaching around 80 $\mu$$\Omega$-cm mol$^2$ K$^2$ J$^{-2}$ for $x$ = 0.3 sample. This increase of KWR by around four times with 30\% of Ti is quite noteworthy which underlines that SrRuO$_3$ has reasonable $U$ which further enhances with introduction of 3$d$ element in present SrRu$_{1-x}$Ti$_x$O$_3$ series. This behavior is consistent with our previous report where the constant $T_c$ in present series has been explained due to simultaneous increase of $U$ and decrease of density of states at Fermi level with substitution of Ti$^{4+}$.\cite{gupta1}
		
\section{Conclusion}
To understand the evolution of electronic correlation and charge transport behavior, a series of polycrystalline SrRu$_{1-x}$Ti$_x$O$_3$ samples are prepared. The parent SrRuO$_3$ shows a metallic behavior throughout the temperature range, however, a distinct correlation between charge conduction and spin ordering is evident below $T_c$. The electrical resistivity of SrRu$_{1-x}$Ti$_x$O$_3$ series increases and a metal to insulator transition has been observed around 40\% of Ti substitution. The electrical resistivity both in PM-metallic state ($x$ $\leq$ 0.4) as well as in insulating state ($x$ $>$ 0.4) follows a modified Mott's VRH model. In FM metallic region, a crossover from $T^2$ to $T^{3/2}$ dependence in resistivity has been observed at low temperature where with applied magnetic field the $T^{3/2}$ dependence prevails over entire FM state. The Fermi liquid and electron-magnon scattering are believed to be reason for this behavior. At low temperature, Kondo like behavior found for samples up to $x$ $\le$ 0.4. The negative MR shows a linear and quadratic increase with magnetic field in FM and PM regime, respectively. Further, a reasonable hysteresis and a crossover from negative to positive value in MR have been observed at low temperature which decrease both with increasing temperature and substitution concentration. This evolution of MR has been associated with the magnetic ordering in the materials. The decreasing value of electronic coefficient of specific heat are in agreement with increasing insulating behavior in present series. We observe a relatively high value of Kadowaki-Woods ratio for SrRuO$_3$ which further increases with substitution indicating an increase of electronic correlation effect with Ti substitution.

\section{Acknowledgment}   
We acknowledge UGC-DAE CSR, Indore for the electrical transport and magnetization measurements. We thank Mr. Kranti Kumar and Mr. Sachin Dabaray for helps in measurements. We are thankful to DST-PURSE for the financial support. RG acknowledges UGC-BSR fellowship.

\end{document}